# Table-Top Milliwatt-Class Extreme Ultraviolet High Harmonic Light Source

R. KLAS[1,2,*], S. DEMMLER[1,2], M. TSCHERNAJEW[1,2], S. HÄDRICH[1,2], Y. SHAMIR[1], A. TÜNNERMANN[1,2,3], J. ROTHHARDT[1,2], J. LIMPERT[1,2,3]

[1]Institute of Applied Physics, Abbe Center of Photonics, Friedrich-Schiller-Universität Jena, Albert-Einstein-Straße 15, 07745 Jena, Germany
[2]Helmholtz Institute Jena, Fröbelstieg 3, 07743 Jena, Germany
[3]Fraunhofer Institute for Applied Optics and Precision Engineering, Albert-Einstein-Straße 7, 07745 Jena, Germany
*Corresponding author: robert.klas@uni-jena.de

**Abstract**
Extreme ultraviolet (XUV) lasers are essential for the investigation of fundamental physics. Especially high repetition rate, high photon flux sources are of major interest for reducing acquisition times and improving signal to noise ratios in a plethora of applications. Here, an XUV source based on cascaded frequency conversion is presented, which delivers due to the drastic better single atom response for short wavelength drivers, an average output power of $(832 \pm 204)$ µW at 21.7 eV. This is the highest average power produced by any HHG source in this spectral range surpassing precious demonstrations by more than a factor of four. Furthermore, a narrow-band harmonic at 26.6 eV with a relative energy bandwidth of only $\Delta E/E = 1.8 \cdot 10^{-3}$ has been generated, which is of high interest for high precision spectroscopy experiments.

Ultrashort extreme ultraviolet (XUV) laser pulses have become one of the most important tools for fundamental studies of atoms and molecules on electronic length- (angstrom) and time-scales (femtosecond to attosecond) [1]. Consequently, they hold great promise to promote scientific and industrial fields covering physics, chemistry, biology, medicine or material science. For that reason, large-scale facilities such as synchrotrons [2] or free electron lasers [3,4] with impressive performance parameters have been developed. However, beam time availability is strongly limited, calling for table-top alternatives. As such, sources based on high harmonic generation (HHG) have gained significant interest over the last decades. In addition to their compact size they offer multiple attractive properties that are not readily available at the large-scale facilities. The HHG process is phase-coupled to the driving laser, thus transferring the laser-like properties, in particular the coherence and short pulse duration, to the XUV spectral region. Additionally, the driving laser pulses are intrinsically synchronized with the XUV pulses, easily enabling pump-probe experiments. As a consequence, such sources have found a plethora of applications. For example, they deliver photon energies suitable for the study of core level transitions, highly exited states, and photoionization or dipole transitions of highly charged ions [5]. Their short wavelengths also allows for imaging techniques with highest spatial resolution [6]. However, in contrast to attosecond physics [7], where a broadband spectrum is desirable for the generation of ultrashort pulses, the above mentioned applications usually require narrow spectral lines. For example, the bandwidth of the employed XUV radiation limits the spatial resolution for coherent diffractive imaging [8], as well as the energy resolution for laser spectroscopy experiments. In particular, material studies, like angle resolved photoelectron spectroscopy, require a relative energy bandwidth of $\Delta E/E = 10^{-2}$ to $10^{-3}$ to resolve the band structure of the investigated materials [9,10]. High precision spectroscopy needs even smaller relative energy bandwidths of $\Delta E/E = 10^{-4}$, e.g. for the test of fundamental theories [5]. In addition, all of the afore-mentioned applications desire a high photon flux for sufficient count rates, statistics, and signal to noise ratios. Especially time-resolved measurements [10] or multi-dimensional studies [11] will benefit from high average power and repetition rates, because much shorter acquisition times get feasible.

To achieve these high average XUV powers either enhancement cavities, exploiting their high intra-cavity average powers [12], or a single pass geometry with a high average power driving laser [13] are commonly used. Since high harmonics are most efficiently generated with ultrashort pulsed driving lasers [14], both requirements of high photon flux and small relative energy bandwidth can usually not be fulfilled simultaneously. However, recently it has been shown that short wavelength drivers allow for very efficient HHG in combination with a narrow energy bandwidth [15,16].

Here we present an XUV source, based on a single pass approach, with a record high photon flux of $(832 \pm 204)$ µW at 21.7 eV, utilizing a fiber based high average power short wavelength driver. At the same time this source provides narrowband harmonics with a relative energy bandwidth of only $\Delta E/E = 8 \cdot 10^{-3}$. By additionally exploiting a Fano-resonance in the absorption spectrum of argon, extremely narrowband ($\Delta E/E = 1.8 \cdot 10^{-3}$) harmonics still carrying µW level average power have been generated.

The experimental setup is shown in Fig. 1. The driver for the system is a fiber chirped pulse amplifier (FCPA) incorporating two coherently combined main amplifier fibers similar to the system presented in [17]. The FCPA is operated with 1 mJ, 300 fs pulses at a repetition rate of 120 kHz and a central wavelength of 1030 nm, which corresponds to 120 W of average power. Afterwards, these pulses are compressed to 45 fs using a nonlinear compression stage based on an argon filled hollow core fiber [18]. A transmission of 55% through this fiber leads to a pulse energy of 0.55 mJ, resulting in 66 W of average power. Subsequently, these pulses are frequency doubled with a 0.5 mm thick beta barium borate (BBO) crystal, delivering a maximum of 11 W of green light with a central wavelength of 515 nm and a pulse duration of 85 fs. The co-propagating fundamental and second harmonic beam are separated using two dichroic mirrors.

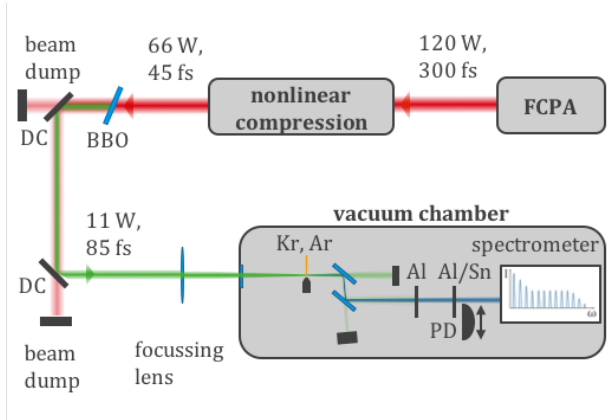

Fig. 1 Experimantal setup. The infrared pulses from the fiber based chirped pulse amplifier (FCPA) are nonlinearly compressed and subsequently frequency doubled in a beta barium borate crystal (BBO). Afterwards, these green pulses are separated from the infrared pulses via dichroic mirrors (DC) and subsequently focused into an argon (Ar) or krypton (Kr) gas jet for high harmonic generation. To separate the generated extreme ultraviolet (XUV) light from the driving laser two fused silica plates in Brewster's angle and thin metal filters of aluminum (Al) or tin (Sn) are used. Finally, the XUV light is analyzed with a XUV photodiode (PD) or a spectrometer.

To generate the XUV light, the 11 W average power second harmonic beam is sent into a vacuum chamber where it is focused to a 35 µm diameter spot size. A 180 µm diameter cylindrical nozzle provides the gas target for the process of HHG. The generated XUV radiation is subsequently separated from the fundamental radiation using two reflections on fused silica plates placed at Brewster's angle for 515 nm. Two additional thin film metal filters, either a 200 nm thick aluminum (Al) and a 200 nm thick tin (Sn) filter, or two 200 nm thick Al filters are used to block the remaining green light and to attenuate the XUV light, in order to avoid saturation of the CCD camera. A flat field spectrometer or a photodiode is used to characterize the XUV radiation.

The photon flux of the higher order harmonics is maximized by iteratively optimizing the target gas density, the position of the gas jet, as well as the opening diameter of an iris in front of the focusing lens. The optimal phase-matching conditions are found via pressure scans revealing an optimum backing pressure of 2.5 bar for argon and 1.5 bar for krypton (red line in Fig. 2a)). The results of a corresponding simulation with a simple one-dimensional model [19,20], shown as the blue line in Fig. 2a), support these findings, assuming a distance between the laser beam and the gas nozzle opening of $6.5w_0$ (with $w_0$ being the $1/e^2$ beam radius) [21]. This results in a pressure of 0.2 bar in the HHG interaction region for argon and 0.3 bar for krypton. Thus, the target gas density ρ can be approximated to be $8.6·10^{18}$ Atoms/cm$^3$ for argon and $5.1·10^{18}$ Atoms/cm$^3$ for krypton. By using the tabulated absorption cross sections σ for both gasses [22] the absorption length can be calculated as $l_{abs}=1/\sigma\rho$ resulting in an absorption length of 32 µm in the case of argon and 51 µm in the case of krypton. In consequence, the medium length $l_{med}$, defined by the diameter of the nozzle (d=180 µm), allows for absorption limited generation of high harmonic radiation ($l_{med} \geq 3 \cdot l_{abs}$) for both generation gases [19]. In addition, our simulations indicate that transient phase-matching could be achieved [19,21], resulting in a high conversion efficiency.

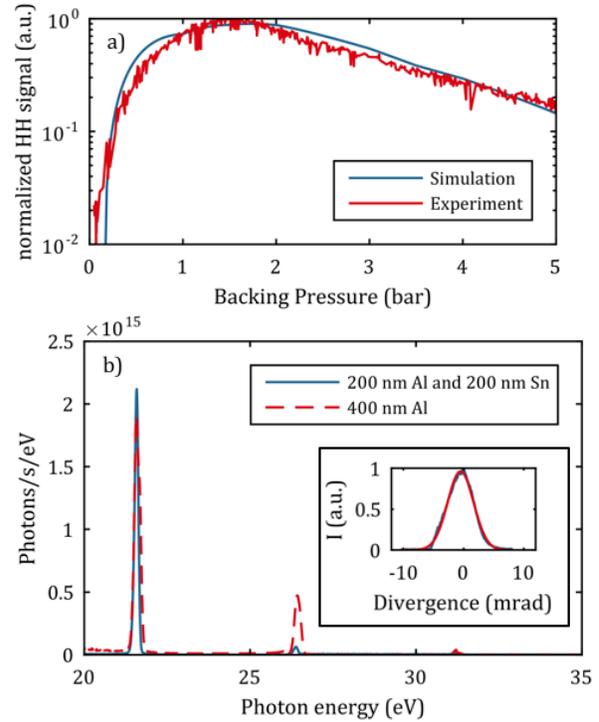

Fig. 2 a) Pressure scan (red) and the corresponding result of our simulation (blue) displaying the intensity of the 9$^{th}$ harmonic versus the applied backing pressure. a) Corresponding spectrum recorded with krypton as generating gas. The red line represents a spectrum measured with two aluminum filters. The blue line displays a spectrum measured with a Sn filter and a Al filter for a cross checking measurement of the power in the 9$^{th}$ harmonic with a photodiode. The inset shows a spatial lineout of the 9$^{th}$ harmonic.

Using krypton as a generating gas yields the highest XUV average power. In order to measure the average power contained in the 9$^{th}$ harmonic only, a Sn foil (which strongly attenuates the adjacent 7$^{th}$ and 11$^{th}$ harmonic (Fig. 2b)) is used. To obtain the XUV photon flux in this harmonic, the source power was determined with a XUV photodiode (calibrated at the National Metrology Institute of Germany (PTB)), taking into account the measured filter transmissions, the reflectivity of the fused silica plates (which also have been calibrated at the PTB), as well as the reabsorption of the gas medium. Note that this reabsorption has been minimized by the use of three turbomolecular pumps, resulting in a background pressure of $1.2·10^{-2}$ mbar in the HHG chamber and $10^{-5}$ mbar in the spectrometer, when a krypton backing pressure of 1.5 bar was applied. This results in a transmission as high as 71 % up to the detector. With these corrections a record high average power of $(832 \pm 204)$ µW has been measured with the photodiode. This is approximately a factor of six higher than previous results obtained with an infrared driving laser in single-pass HHG [23] and a factor of 4 higher than what has been achieved with enhancement cavities [12]. The resulting conversion efficiency of green to XUV light is as high as $7.5·10^{-5}$, and the total conversion efficiency from the infrared to XUV is $1.2·10^{-5}$, which is an order of magnitude improvement compared to HHG directly from an infrared driving laser [23]. This high conversion efficiency is a result of a longer phase matching time window [16], compared to infrared pulses, in combination with the $\sim\lambda^{-5}$ scaling [24] of the single atom response.

In addition to the XUV photodiode, a flat-field spectrometer equipped with a CCD camera was used to record the harmonic spectrum and to estimate the average power, taking into account the known detection efficiencies and quantum efficiencies of the employed CCD camera (supplied by the manufacturer), the measured filter transmissions, the reabsorption due to the remaining gas in the vacuum chamber, as well as the reflectivities of the calibrated fused silica plates. Furthermore, spatial cutting of the XUV beam due to the limited size of the grating and the detector is taken into account. The resulting spectrum is shown in Fig. 2b) and the average power in the 9th harmonic at 21.7 eV is $(1.5 \pm 0.6)$ mW, which corresponds to $(4.3 \pm 1.7) \cdot 10^{14}$ photons per second. Within the error intervals this evaluation confirms the record high photon flux, which has been measured with the XUV photodiode.

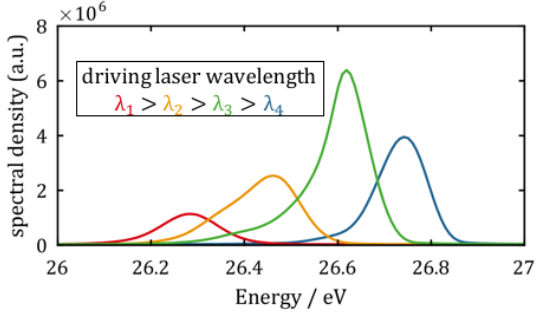

Fig. 3 Tunability of the 11th harmonic (generated in argon) due to different wavelengths of the driving laser, achieved by changing the phase-matching angle of the SHG-crystal.

This XUV light source does not only provide an exceptionally high photon flux., but due to the cascaded frequency conversion via SHG followed by HHG, the central wavelength of the driving laser can be tuned. Hence, the spectral position of the higher order harmonics is tunable in a certain spectral range of ~0.5 eV (Fig. 3). This wavelength tuning of the driving laser is simply done by changing the phase matching angle of the BBO. The BBO thickness is chosen slightly thicker than required to achieve phase-matched SHG for the entire bandwidth of the infrared laser. Therefore, turning of the crystal leads to frequency doubling of different spectral portions and, consequently, central wavelengths. Due to the associated changes in the SHG power this results in a slightly lower harmonic power, when tuning off-center, as shown in Fig. 3.

This wavelength tuning together with the unique properties of argon opens an extremely interesting feature for high precision spectroscopy at 26.65 eV [25]. At this energy, argon has a window-type Fano-resonance in the absorption spectrum which results in a much higher transmission through the gas medium than at other photon energies [26]. As a first experiment krypton is used as a generation gas for HHG and the maximum SHG average power is used. The 25resulting spectrum of the 11th harmonic, shown in Fig. 4a), reveals a relative linewidth of $\Delta E/E = 8 \cdot 10^{-3}$ and an average power of 360 µW. The linewidth of this harmonic can be considered as a typical linewidth depending on the pulse duration and the chirp of the driving laser. Compared to the linewidth of harmonics generated directly with an infrared driving laser at similar conditions [23], this typical linewidth is already a factor of three smaller, while containing more than twice as much average power. If a narrower linewidth is desired, argon can be used as a generating gas. For exploiting its resonance at 26.65 eV the energy of the 11th harmonic needs to be shifted, as previously explained. Using the optimal backing pressure of 2.9 bar results in an average power of 72 µW and a relative energy bandwidth of $\Delta E/E = 4.2 \cdot 10^{-3}$ (Fig. 4b)), which is an improvement of a factor of two compared to the krypton harmonic in Fig. 4a). Since the reabsorption of the XUV light mostly takes place in the HHG interaction region [27] the backing pressure needs to be increased in order to achieve even narrower bandwidths with this experimental setup. By increasing the backing pressure to the maximum pressure manageable by the turbomolecular pumps (6.5 bar), an average power of 6 µW and a relative bandwidth of only $\Delta E/E = 8 \cdot 10^{-3}$ was achieved, which is at the resolution limit of the used spectrometer (see Fig. 4c)). Note that in this case all other harmonics are suppressed by at least one order of magnitude, due to the strong reabsorption of the XUV light in argon. So there is no need for additional filters. Moreover, the spectral position of the incorporated resonance is well known [26] and intrinsic to the generating gas. Although, since the conversion efficiency is reduced in this case, such a configuration is highly interesting for precision spectroscopy experiments.

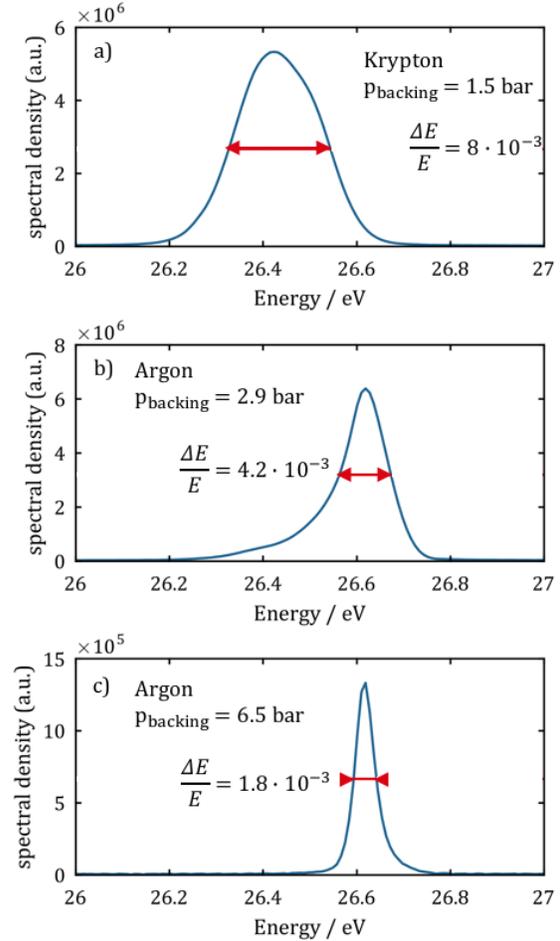

Fig. 4 Linewidth of the 11th harmonic at ~26.6 eV generated with different media at different backing pressures. a) harmonic generated in krypton with an average power of $\sim 360$ µW, b) harmonic generated in argon at optimized phase matching pressure with $\sim 72$ µW, and c) harmonic generated in argon at high backing pressure and $\sim 6$ µW.

In conclusion, a narrowband high photon flux HHG source has been presented. It is driven by a nonlinearly compressed and frequency doubled fiber based CPA, delivering 90 µJ pulses with a pulse duration of 85 fs at 515 nm central wavelength and an average power of 11 W. The unique combination of a high average power driving laser and high conversion efficiency due to the short driving wavelength results in a record high XUV average power of $(832 \pm 204)$ µW $((2.4 \pm 0.6) \cdot 10^{14}$ photons/s) in the 9th harmonic at 21.7 eV. This table-top source provides the highest average power in this spectral region available to date, even delivering a higher photon flux than femto-sliced synchrotrons [2]. In particular, photon hungry applications like time resolved coincidence measurements [17] or imaging of ultrafast dynamics [28], which additionally need ultrashort pulses, will benefit from such a source.

Furthermore, a strong window-type Fano-resonance in argon allowed to reduce the relative energy bandwidth of a single harmonic at 26.65 eV to the resolution limit of the used spectrometer of only $\Delta E/E = 8\cdot10^{-3}$. Nevertheless, some applications, like high resolution imaging and precision spectroscopy might demand for even smaller bandwidths of $\Delta E/E = 10^{-4}$. This seems feasible by using driving pulses in the ultraviolet spectral region [29]. Furthermore, the coherently combined fiber based driving laser is average power scalable [30], which promises an order of magnitude higher photon flux in the near future.

**FUNDING SOURCES AND ACKNOWLEDGMENTS**

We acknowledge support by the German Ministry of Education and Research(BMBF) under project number 05P2015 (APPA R&D: Licht-Materie Wechselwirkung mit hochgeladenen Ionen).

**REFERENCES**

1. F. Krausz and M. Ivanov, *Rev. Mod. Phys.*, vol. 81, no. 1, 163–234 (2009).
2. S. Khan, K. Holldack, T. Kachel, R. Mitzner, and T. Quast, *Phys. Rev. Lett.*, vol. 97, no. 7, 1–4 (2006).
3. J. D. Bozek, *Eur. Phys. J. Spec. Top.*, vol. 169, no. 1, 129–132 (2009).
4. V. Ayvazyan, N. Baboi, J. Bähr, V. Balandin, B. Beutner, A. Brandt, I. Bohnet, A. Bolzmann, R. Brinkmann, O. I. Brovko, J. P. Carneiro, S. Casalbuoni, M. Castellano, P. Castro, L. Catani, E. Chiadroni, S. Choroba, A. Cianchi, H. Delsim-Hashemi, G. di Pirro, M. Dohlus, S. Düsterer, H. T. Edwards, B. Faatz, A. A. Fateev, J. Feldhaus, K. Flöttmann, J. Frisch, L. Fröhlich, T. Garvey, U. Gensch, N. Golubeva, H. J. Grabosch, B. Grigoryan, O. Grimm, U. Hahn, J. H. Han, M. V. Hartrott, K. Honkavaara, M. Hüning, R. Ischebeck, E. Jaeschke, M. Jablonka, R. Kammering, V. Katalev, B. Keitel, S. Khodyachykh, Y. Kim, V. Kocharyan, M. Körfer, M. Kollewe, D. Kostin, D. Krämer, M. Krassilnikov, G. Kube, L. Lilje, T. Limberg, D. Lipka, F. Löhl, M. Luong, C. Magne, J. Menzel, P. Michelato, V. Miltchev, M. Minty, W. D. Möller, L. Monaco, W. Müller, M. Nagl, O. Napoly, P. Nicolosi, M. Nölle, T. Nuñez, A. Oppelt, C. Pagani, R. Paparella, B. Petersen, B. Petrosyan, J. Pflüger, P. Piot, E. Plönjes, L. Poletto, D. Proch, D. Pugachov, K. Rehlich, D. Richter, S. Riemann, M. Ross, J. Rossbach, M. Sachwitz, E. L. Saldin, W. Sandner, H. Schlarb, B. Schmidt, M. Schmitz, P. Schmüser, J. R. Schneider, E. A. Schneidmiller, H. J. Schreiber, S. Schreiber, A. V. Shabunov, D. Sertore, S. Setzer, S. Simrock, E. Sombrowski, L. Staykov, B. Steffen, F. Stephan, F. Stulle, K. P. Sytchev, H. Thom, K. Tiedtke, M. Tischer, R. Treusch, D. Trines, I. Tsakov, A. Vardanyan, R. Wanzenberg, T. Weiland, H. Weise, M. Wendt, I. Will, A. Winter, K. Wittenburg, M. V. Yurkov, I. Zagorodnov, P. Zambolin, and K. Zapfe, *Eur. Phys. J. D*, vol. 37, no. 2, 297–303 (2006).
5. J. Rothhardt, S. Hädrich, S. Demmler, M. Krebs, D. Winters, T. Kuehl, T. Stöhlker, J. Limpert, and A. Tünnermann, Phys. Scr. **T166**, 14030 (2015).
6. G.K. Tadesse, R. Klas, S. Demmler, S. Hädrich, I. Wahyutama, M. Steinert, C. Spielmann, M. Zürch, A. TÜnnermann, J. Limpert, J. Rothhardt, arXiv preprint arXiv:1605.02909 (2016).
7. F. Krausz and M. Ivanov, "Attosecond physics," Rev. Mod. Phys. **81**, 163–234 (2009).
8. M. Zürch, J. Rothhardt, S. Hädrich, S. Demmler, M. Krebs, J. Limpert, a. Tünnermann, a. Guggenmos, U. Kleineberg, and C. Spielmann, Sci. Rep. **4**, 7356 (2014).
9. A. Damascelli, Z. Hussain, and Z. X. Shen, Rev. Mod. Phys. **75**, 473 (2003).
10. T. Rohwer, S. Hellmann, M. Wiesenmayer, C. Sohrt, A. Stange, B. Slomski, A. Carr, Y. Liu, L. M. Avila, M. Kalläne, S. Mathias, L. Kipp, K. Rossnagel, and M. Bauer, *Nature*, vol. 471, no. 7339, 490–493 (2011).
11. S. Mukamel, D. Healion, Y. Zhang, and J. D. Biggs, *Annu. Rev. Phys. Chem.*, vol. 64, no. March, pp. 101–27, 2013.
12. A. Cingöz, D. C. Yost, T. K. Allison, A. Ruehl, M. E. Fermann, I. Hartl, and J. Ye, *Nature*, vol. 482, no. 7383, 68–71 (Feb. 2012).
13. S. Hädrich, J. Rothhardt, M. Krebs, S. Demmler, A. Klenke, A. Tünnermann, and J. Limpert, J. Phys. B., accepted.
14. K. J. Schafer and K. C. Kulander, *Phys. Rev. Lett.*, vol. 78, no. 4, pp. 638–641, 1997.
15. D. Popmintchev, C. Hernandez-Garcia, F. Dollar, C. Mancuso, J. A. Perez-Hernandez, M.-C. Chen, A. Hankla, X. Gao, B. Shim, A. L. Gaeta, M. Tarazkar, D. A. Romanov, R. J. Levis, J. A. Gaffney, M. Foord, S. B. Libby, A. Jaron-Becker, A. Becker, L. Plaj a, M. M. Murnane, H. C. Kapteyn, and T. Popmintchev, *Science*, vol. 350, no. 6265, 1225–1231 (2015).
16. H. Wang, Y. Xu, S. Ulonska, P. Ranitovic, J. S. Robinson, and R. A. Kaindl, Nat. Comm 6:7459, 1–15 (2015).
17. Jan Rothhardt, Steffen Hädrich, Yariv Shamir, Maxim Tschnernajew, Robert Klas, Armin Hoffmann, Getnet K. Tadesse, Arno Klenke, Thomas Gottschall, Tino Eidam, Jens Limpert, Andreas Tünnermann, Rebecca Boll, Cedric Bomme, Hatem Dachraoui, Benjamin Erk, Michele Di Fraia, Daniel A. Horke, Thomas Kierspel, Terence Mullins, Andreas Przystawik, Evgeny Savelyev, Joss Wiese, Tim Laarmann, Jochen Küpper, Daniel Rolles, arXiv preprint arXiv:1602.03703 (2016).
18. S. Hädrich, J. Rothhardt, T. Eidam, J. Limpert, and A. Tünnermann, *Opt. Express*, vol. 17, no. 5, 3913–3922 (2009).
19. E. Constant, D. Garzella, P. Breger, E. Mével, C. Dorrer, C. Le Blanc, F. Salin, and P. Agostini, *Phys. Rev. Lett.*, vol. 82, no. 8, 1668–1671 (1999).
20. S. Kazamias, S. Daboussi, O. Guilbaud, K. Cassou, D. Ros, B. Cros, and G. Maynard, *Phys. Rev. A*, vol. 83, 1–6 (2011).
21. J. Rothhardt, M. Krebs, S. Hädrich, S. Demmler, J. Limpert and A. Tünnermann, New J. of Phys., Vol. 16, Iss 3, 033022 (2014).
22. J. A. R. Samson and W. C. Stolte, vol. 123, pp. 265–276 (2002).
23. S. Hädrich, A. Klenke, and J. Rothhardt, *Nat. Photonics*, vol. 8, no. 10, 780 (2014).
24. M. V. Frolov, N. L. Manakov, and A. F. Starace, *Phys. Rev. Lett.*, vol. 100, no. 17, 1–4 (2008).
25. H. Kjeldsen, F. Folkmann, J. E. Hansen, H. Knudsen, M. S. Rasmussen, J. B. West, and T. Andersen, *Astrophys. J. Lett.*, vol. 524, no. 2, 3–6 (1999).
26. R. P. Madden, D. L. Ederer, and K. Codling, Phys. Rev. **177**, 136–151 (1969).
27. J. Rothhardt, S. Hädrich, S. Demmler, M. Krebs, S. Fritzsche, J. Limpert, and A. Tünnermann, *Phys. Rev. Lett.*, vol. 112, no. 23, 1–5 (2014).
28. J. Miao, T. Ishikawa, I. K. Robinson, and M. Murnane, *Science*, vol. 348, no. 6234, 530–535 (2015).
29. J. Rothhardt, C. Rothhardt, M. Müller, A. Klenke, M. Kienel, S. Demmler, T. Elsmann, M. Rothhardt, J. Limpert, and A. Tünnermann, *Opt. Lett.*, vol. 41, no. 8, 1885 (2016).
30. M. Müller, M. Kienel, A. Klenke, T. Gottschall, E. Shestaev, M. Plötner, J. Limpert, and A. Tünnermann, Opt. Lett. submitted (n.d.).